\documentstyle[floats,twocolumn,epsfig,amssymb,prl,aps]{revtex}

\begin{document}

\draft

\wideabs{%
\title{Triangular Trimers on the Triangular Lattice: an Exact Solution}
\author{Alain Verberkmoes \and Bernard Nienhuis}
\address{Instituut voor Theoretische Fysica, Universiteit van
  Amsterdam, Valckenierstraat 65, 1018 XE Amsterdam, The Netherlands}
\date{April 23, 1999}

\maketitle

\begin{abstract}%
\noindent
A model is presented consisting of triangular trimers on the triangular
lattice.  In analogy to the dimer problem, these particles cover the
lattice completely without overlap.  The model has a honeycomb
structure of hexagonal cells separated by rigid domain walls.  The
transfer matrix can be diagonalised by a Bethe Ansatz with two types of
particles.  This leads two an exact expression for the entropy on a
two-dimensional subset of the parameter space.
\end{abstract}
\pacs{PACS number(s): 05.50.+q}
}

In the course of years a few exactly solvable lattice gas models have
been found.  The Ising model~\cite{ising:1925}, proposed in 1920 by
Lenz~\cite{lenz:1920} as a model of a ferromagnet, can be interpreted
as a lattice gas with hard-core repulsion and short-range attraction.
The (zero-field square lattice) Ising model was solved in 1944 by
Onsager~\cite{onsager:1944}.  It exhibits gas--liquid coexistence below
a critical temperature and a single fluid phase above.

Another lattice gas is the hard hexagon model~\cite{burley:1960},
solved in 1980 by Baxter~\cite{baxter:1980}.  It has a continuous
fluid--solid transition.  At high-density (solid) the particles select
one of three sub-lattices; at low-density (fluid) these sub-lattices
are evenly occupied.

As a final example we mention the dimer problem.  It was solved for
planar lattices in 1961, independently by
Kasteleyn~\cite{kasteleyn:1961} and by Temperley and
Fisher~\cite{temperley:1961}.  A dimer is a particle that occupies two
adjacent lattice sites.  As in the Ising and the hard hexagon model two
particles cannot occupy the same lattice site.  In contrast to these
models it is also required that all sites are occupied.  The
configurations are coverings of a lattice with dimers, without empty
sites or overlap.  The dimer problem is reviewed in
Ref.~\cite{nagle:1989}.  We discuss the dimer model on the honeycomb
lattice in some detail, because it has illustrative similarities to a
new model we shall introduce below.

\begin{figure}[!t]
  \hfil\epsfig{file=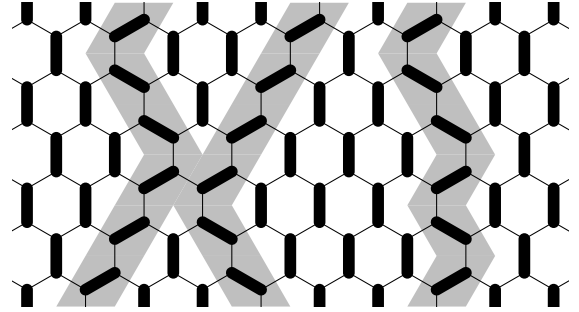,width=0.9 \columnwidth}\hfil
  \caption{The dimer model on the honeycomb lattice.  Domains of
    vertical dimers are separated by domain walls consisting of dimers
    of the other two orientations.  To guide the eye the domain walls
    are shaded.}
  \label{fig:dimers}
\end{figure}

A configuration of the honeycomb lattice dimer model can be viewed as a
number of domains consisting of vertical dimers, separated by
zigzagging domain walls made up of dimers of the other two
orientations.  This is illustrated in Fig.~\ref{fig:dimers}.  The
domain walls run from the bottom to the top of the lattice, so that any
horizontal line through the system meets all domain walls.  Hence the
number of domain walls is the same in each horizontal slice, in other
words, it is a conserved quantity.

Consider the entropy for fixed density $\rho$ of domain wall dimers.
From the exact solution of the model it can be calculated that for low
$\rho$ the entropy per dimer is given by
\begin{displaymath}
  S \approx (\log 2) \rho - \frac{\pi^2}{24} \rho^3.
\end{displaymath}
The linear term reflects the zigzag freedom of the domain walls; each
domain wall dimer contributes $\log 2$ to the entropy.  The cubic term
is due to the (repulsive) interaction between the domain walls: when
two domain walls meet some of the zigzag freedom is lost.

Now give chemical potentials $\mu$ to the dimers in the domain walls
and $0$ to the vertical dimers.  For $\mu \le -\log 2$ the free energy
$F = -\mu \rho - S(\rho)$ is is an increasing function of $\rho$ for
small $\rho$, so no domain walls will be present.  For
$\mu \gtrsim -\log 2$ the free energy has a minimum at some small
positive value of~$\rho$.  At $\mu = -\log 2$ there is a transition
between a frozen phase consisting of vertical dimers only and a rough
phase where dimers of all three orientations are present.

\begin{figure}[!t]
  \hfil\epsfig{file=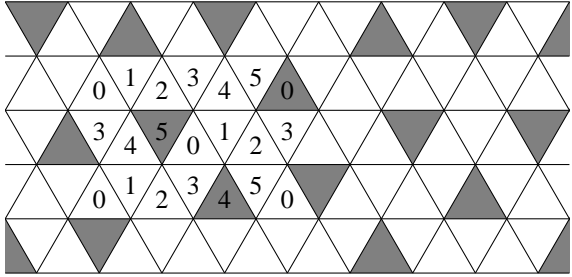,width=0.9 \columnwidth}\hfil
  \caption{A typical configuration of the trimer model.  Each lattice
    site belongs to precisely one trimer.  We divide the lattice faces
    into six sub-lattices, numbered 0, 1, \dots,~5.  Filling one
    sub-lattice completely and leaving the other five empty gives a
    very regular configuration of the model.}
  \label{fig:trimers}
\end{figure}

Inspired by the dimer model we consider coverings of the triangular
lattice by triangular trimers.  A trimer is a particle that occupies
three lattice sites.  As in the dimer problem we require that there are
no empty sites and that there is no overlap.  Fig.~\ref{fig:trimers}
shows a typical configuration.  In this paper we present our main
results on this model.  We intend to publish a more detailed account
later.

The model admits very regular configurations where the trimers occupy
a sub-lattice of the triangular faces.  There are six such
sub-lattices, which we number 0, 1, \dots, 5 as indicated in
Fig.~\ref{fig:trimers}.  Note that the up and down triangles make up
the even-numbered and odd-numbered sub-lattices, respectively.

\begin{figure}[!b]
  \hfil\epsfig{file=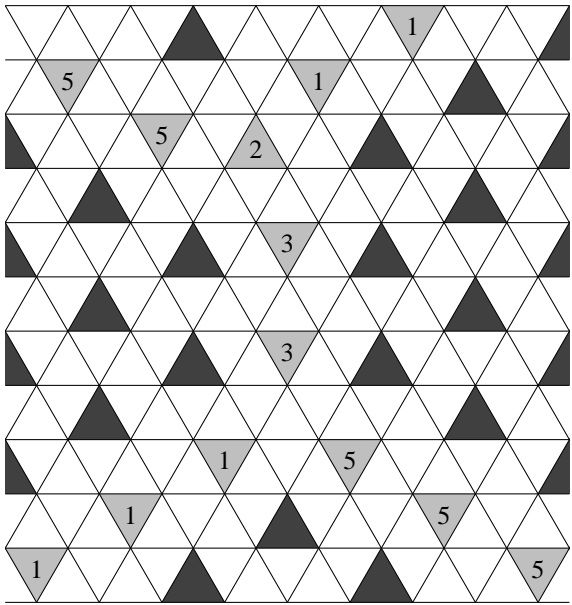,width=0.9 \columnwidth}\hfil
  \caption{The trimer model has three types of domain walls,
    corresponding to the three odd-numbered sub-lattices.  The trimers
    on sub-lattice 0 are shaded dark grey; the trimers on the other
    sub-lattices are shaded light grey, and their sub-lattice number is
    given.}
  \label{fig:coll}
\end{figure}

Consider configurations where trimers on sub-lattice 0 predominate.
They consist of hexagonal domains of trimers on this sub-lattice,
separated by straight domain walls that form an irregular honeycomb
network.  There are three types of domain walls, of different
orientations.  The domain walls that run from lower right to upper left
will be termed~L; they are made up of trimers on sub-lattice~5.  Those
running from lower left to upper right will be called~R; they consist
of trimers on sub-lattice~1.  The vertical domain walls are made up of
trimers on sub-lattice~3.  When domain walls of the three different
types meet in a Y-shape a trimer on sub-lattice 2 or 4 occurs.  This is
illustrated in Fig.~\ref{fig:coll}.

For the time being we require that the model is isotropic, in the sense
that there are equal amounts of the three types of domain walls.  Let
$\rho$ denote the density of domain wall trimers.  The domain walls are
rigid, so they have no zigzag freedom contributing to the entropy.
Therefore the low density expansion of the entropy contains no term
linear in $\rho$.  There is, however, freedom in the sizes of the
domains~\cite{villain:1980}.  For example, it is possible to enlarge a
single domain while simultaneously shrinking its six neighbours.  The
contribution per domain depends on the linear dimensions of the
domains, and is roughly proportional to~$-\log \rho$.  The number of
domains is approximately proportional to~$\rho^2$.  Hence the
``breathing'' entropy is given for low $\rho$ by
\begin{equation}
  S \approx -K \rho^2 \log \rho,
  \label{equ:trimentr}
\end{equation}
where $K$ is some (positive) proportionality constant.

If a chemical potential $\mu$ is given to the domain wall trimers, the
free energy for low $\rho$ is
\begin{displaymath}
  F \approx -\mu \rho + K \rho^2 \log \rho.
\end{displaymath}
This is an increasing function of $\rho$ for $\mu < 0$ and a decreasing
function for~$\mu \ge 0$.  Hence the free energy takes its minimum
either at $\rho = 0$ or at a large value of $\rho$, for which the
approximation (\ref{equ:trimentr}) is not valid.  For small $\mu$ there
are no domain walls, but when $\mu$ passes some threshold $\rho$ jumps
to a positive value.  Thus the phase transition is different from that
in the honeycomb lattice dimer model, where the domain wall density
increases gradually at the phase transition.

Now we return to the model without the isotropy requirement.  Let $N$
denote the total number of trimers, $N_i$ the number of trimers on
sub-lattice $i$, and $\rho_i$ the partial density~$N_i/N$.  These six
sub-lattice densities obviously satisfy
\begin{equation}
  \rho_0 + \rho_1 + \rho_2 + \rho_3 + \rho_4 + \rho_5 = 1.
  \label{equ:lin}
\end{equation}
It can be shown that, when toroidal boundary conditions are imposed,
they also satisfy
\begin{equation}
  \rho_0 \rho_2 + \rho_2 \rho_4 + \rho_4 \rho_0 =
  \rho_1 \rho_3 + \rho_3 \rho_5 + \rho_5 \rho_1.
  \label{equ:qua}
\end{equation}

When the total density of down trimers
$\rho_{\triangledown} = \rho_1 + \rho_3 + \rho_5$ is small, it follows
easily from (\ref{equ:lin}) and (\ref{equ:qua}) that one of $\rho_0$,
$\rho_2$ and $\rho_4$, say $\rho_0$, is larger than the other two.  If
there is no further symmetry breaking
$\rho_0 > \rho_1 = \rho_3 = \rho_5 > \rho_2 = \rho_4$.  By the same
token when $\rho_{\triangledown}$ is close to $1$ the symmetry between
the down sub-lattices is broken.  Therefore when $\rho_{\triangledown}$
is increased from $0$ to $1$ at least one phase transition is expected.

Beside (\ref{equ:lin}) and (\ref{equ:qua}) we have found no more
constraints on the sub-lattice densities.
Therefore of the six sub-lattice densities four are independent.
We would like to know the entropy (per trimer) as function of these
four parameters.  We have been able to compute it for a two-dimensional
subset of the four-dimensional parameter space.  The calculation is
rather lengthy, so here we only give an outline of the method, and a
description of the final result.

View each vertical domain wall as a combination of one L domain wall
and one R domain wall.  Then the L and R domain walls run without
interruption from the bottom to the top of the lattice.  Therefore the
number of L domain walls and the number of R domain walls are constant
throughout the system.  This is analogous to the situation described
above for the dimer model on the honeycomb lattice, except that there
are now two conserved quantities $n_{\text{L}}$ and $n_{\text{R}}$
instead of a single one.  We introduce the densities
$\rho_{\text{L}} = n_{\text{L}}/L$ and
$\rho_{\text{R}} = n_{\text{R}}/L$, where $3L$ is the number of sites
in a horizontal row of the lattices.  They can be expressed in terms of
the sub-lattice densities:
\begin{eqnarray}
  \rho_{\text{L}} &=& 1 - \rho_0 - \rho_1 + \rho_3 + \rho_4, \nonumber
  \\
  \rho_{\text{R}} &=& 1 - \rho_0 + \rho_2 + \rho_3 - \rho_5. \nonumber
\end{eqnarray}
It is suggestive to interpret the vertical lattice direction as
``time'' and the horizontal direction as ``space''.  The domain walls
then are viewed as world lines of two types of particles, L and~R.  In
a vertical domain wall one L and one R particle form a ``bound
state''.

The model can be formulated in terms of a transfer matrix, which
describes the ``time'' evolution of the system of L particles and
R particles in one ``space'' dimension.  Solving the model boils down
to determining the largest eigenvalue of this operator, or more
precisely, its maximum over all particle numbers $n_{\text{L}}$
and~$n_{\text{R}}$.  This we have achieved using coordinate Bethe
Ansatz; the solution is similar to that of the square-triangle random
tiling model, due to Widom~\cite{widom:1993} and
Kalugin~\cite{kalugin:1994}.  A model can be solved in this way only
if, in some sense, the many-particle interactions factorise into
two-particle interactions.  It turns out that for the present model
this is indeed the case.  It is noteworthy that the L particles among
each other are free fermions, as are the R particles, but that the
interaction between an L and an R particle is non-trivial.

The Bethe Ansatz allows for numerical computations for the system on an
infinitely long cylinder of finite circumference.  These computations
can be done to arbitrary precision, and effectively for the full
four-dimensional parameter space of the model.  In the thermodynamic
limit the Bethe Ansatz gives rise to a set of two coupled integral
equations.  The physical quantities we are interested in, such as the
densities $\rho_{\text{L}}$ and $\rho_{\text{R}}$ and the entropy, can
be expressed in terms of the functions satisfying these equations.
These can be solved analytically in a special case~\cite{kalugin:1994}.
Thus we have obtained an exact expression for the entropy for a
two-dimensional family of sub-lattice densities.  It will be seen below
that this family is given by
$\rho_1 = \rho_3 = \rho_5$ (or $\rho_0 = \rho_2 = \rho_4$).

\begin{figure}[!t]
  \hfil\epsfig{file=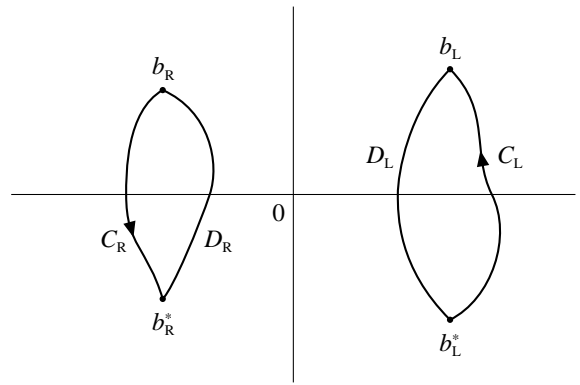,width=0.9 \columnwidth}\hfil
  \caption{The exact expression for the entropy is formulated in terms
    of contour integrals.  The curves $C_{\text{L}}$ and $D_{\text{L}}$
    in the right half plane are deformations of the line segment
    joining $b_{\text{L}}$ and $b_{\text{L}}^*$, and $C_{\text{L}}$
    lies to the right of $D_{\text{L}}$.  No orientation of
    $D_{\text{L}}$ needs to be specified because it does not occur as
    an integration contour but only as a branch cut.  Analogous remarks
    apply to $C_{\text{R}}$ and~$D_{\text{R}}$.}
  \label{fig:contours}
\end{figure}

This solution is parametrised by a complex number $\hat b$ with
$\text{Im } \hat b > 0$.  Write
\begin{equation}
  \hat b =
  b_{\text{L}}-b_{\text{L}}^{-1} = b_{\text{R}}-b_{\text{R}}^{-1}
  \label{equ:bhat}
\end{equation}
with $\text{Re } b_{\text{L}} \ge 0$ and
$\text{Re } b_{\text{R}} \le 0$.  It follows that
$\text{Im } b_{\text{L}} > 0$, $\text{Im } b_{\text{R}} > 0$ and
$b_{\text{L}} b_{\text{R}} = -1$.  Take contours $C_{\text{L}}$ and
$D_{\text{L}}$ running from $b_{\text{L}}^*$ to $b_{\text{L}}$, and
$C_{\text{R}}$ and $D_{\text{R}}$ running from $b_{\text{R}}$
to~$b_{\text{R}}^*$.  The arrangement of these four curves must be as
shown in Fig.~\ref{fig:contours}, but their precise shape is
immaterial.  (In fact there are more solutions, corresponding to
contour configurations different from that in Fig.~\ref{fig:contours}.
Since they are related by permutations of the six sub-lattices, we only
treat one case here.)  Define the complex function
\begin{displaymath}
  t(z) =
  \left( \frac{z-z^{-1}-\hat b}{z-z^{-1}-\hat b^*} \right)^{1/6}.
\end{displaymath}
Fix a branch $t_{\text{L}}(z)$ with branch cuts $C_{\text{R}}$ and
$D_{\text{L}}$ by $t_{\text{L}}(0) = \exp(\pi \text{i}/3)$, and a
branch $t_{\text{R}}(z)$ with branch cuts $C_{\text{L}}$ and
$D_{\text{R}}$ by $t_{\text{R}}(0) = \exp(-\pi \text{i}/3)$.
The domain wall density $\rho_{\text{L}}$ is given by
\begin{displaymath}
  \rho_{\text{L}} =
  \frac{1}{2 \pi \text{i}} \int_{C_{\text{L}}}
  \frac{t_{\text{L}}(z)+t_{\text{L}}(z)^{-1}}{z} \, \text{d} z,
\end{displaymath}
and $\rho_{\text{R}}$ is given by the same equation with all subscripts
L changed into~R.  The sub-lattice densities are
\begin{eqnarray}
  \rho_0 &=&
  1 - \frac{1}{2} (\rho_{\text{L}} + \rho_{\text{R}}) + \frac{1}{6}
  (\rho_{\text{L}}^2 - \rho_{\text{L}} \rho_{\text{R}} +
  \rho_{\text{R}}^2), \nonumber \\
  \rho_2 &=&
  \frac{1}{2} (\rho_{\text{R}} - \rho_{\text{L}}) + \frac{1}{6}
  (\rho_{\text{L}}^2 - \rho_{\text{L}} \rho_{\text{R}} +
  \rho_{\text{R}}^2), \nonumber \\
  \rho_4 &=&
  \frac{1}{2} (\rho_{\text{L}} - \rho_{\text{R}}) + \frac{1}{6}
  (\rho_{\text{L}}^2 - \rho_{\text{L}} \rho_{\text{R}} +
  \rho_{\text{R}}^2), \nonumber \\
  \rho_i &=&
  \frac{1}{6} (\rho_{\text{L}} + \rho_{\text{R}}) - \frac{1}{6}
  (\rho_{\text{L}}^2 - \rho_{\text{L}} \rho_{\text{R}} +
  \rho_{\text{R}}^2) \quad \text{for odd $i$}.
  \label{equ:odd}
\end{eqnarray}
Define auxiliary integrals $\phi_{\text{L}}$ and $\Sigma_{\text{L}}$ by
\begin{eqnarray}
  \phi_{\text{L}} &=&
  \frac{1}{2} \text{Re} \int_{b_{\text{L}}}^{-b_{\text{L}}^{-1}}
  \frac{t_{\text{L}}(z)+t_{\text{L}}(z)^{-1}}{z} \, \text{d} z,
  \nonumber \\
  \Sigma_{\text{L}} &=&
  \frac{1}{4} \text{Re} \int_0^\infty
  \frac{t_{\text{L}}(z)+t_{\text{L}}(z)^{-1}-1}{z} \, \text{d} z.
  \nonumber
\end{eqnarray}
The real parts of these integrals do not depend on the choice of the
integration contours, which must not meet the branch cuts
$C_{\text{R}}$ and~$D_{\text{L}}$.  The auxiliary integrals
$\phi_{\text{R}}$ and $\Sigma_{\text{R}}$ are defined analogously.  The
entropy per trimer is given by
\begin{displaymath}
  S =
  \Sigma_{\text{L}} + \Sigma_{\text{R}} +
  \frac{1}{6} (2 \rho_{\text{R}}-\rho_{\text{L}}) \phi_{\text{L}} +
  \frac{1}{6} (2 \rho_{\text{L}}-\rho_{\text{R}}) \phi_{\text{R}}.
\end{displaymath}

We started out with the model parametrised by the six sub-lattice
densities satisfying the constraints (\ref{equ:lin})
and~(\ref{equ:qua}).  The exact solution described above has two
parameters, so it covers a two-dimensional set of sub-lattice
densities.  The solution is however parametrised by a complex number
$\hat b$, and not in terms of these densities.  It follows from
(\ref{equ:odd}) that $\rho_1 = \rho_3 = \rho_5$ for the exact solution.
The space of sub-lattice densities satisfying this constraint as well
as (\ref{equ:lin}) and (\ref{equ:qua}) is two-dimensional.

\begin{figure}[!t]
  \hfil\epsfig{file=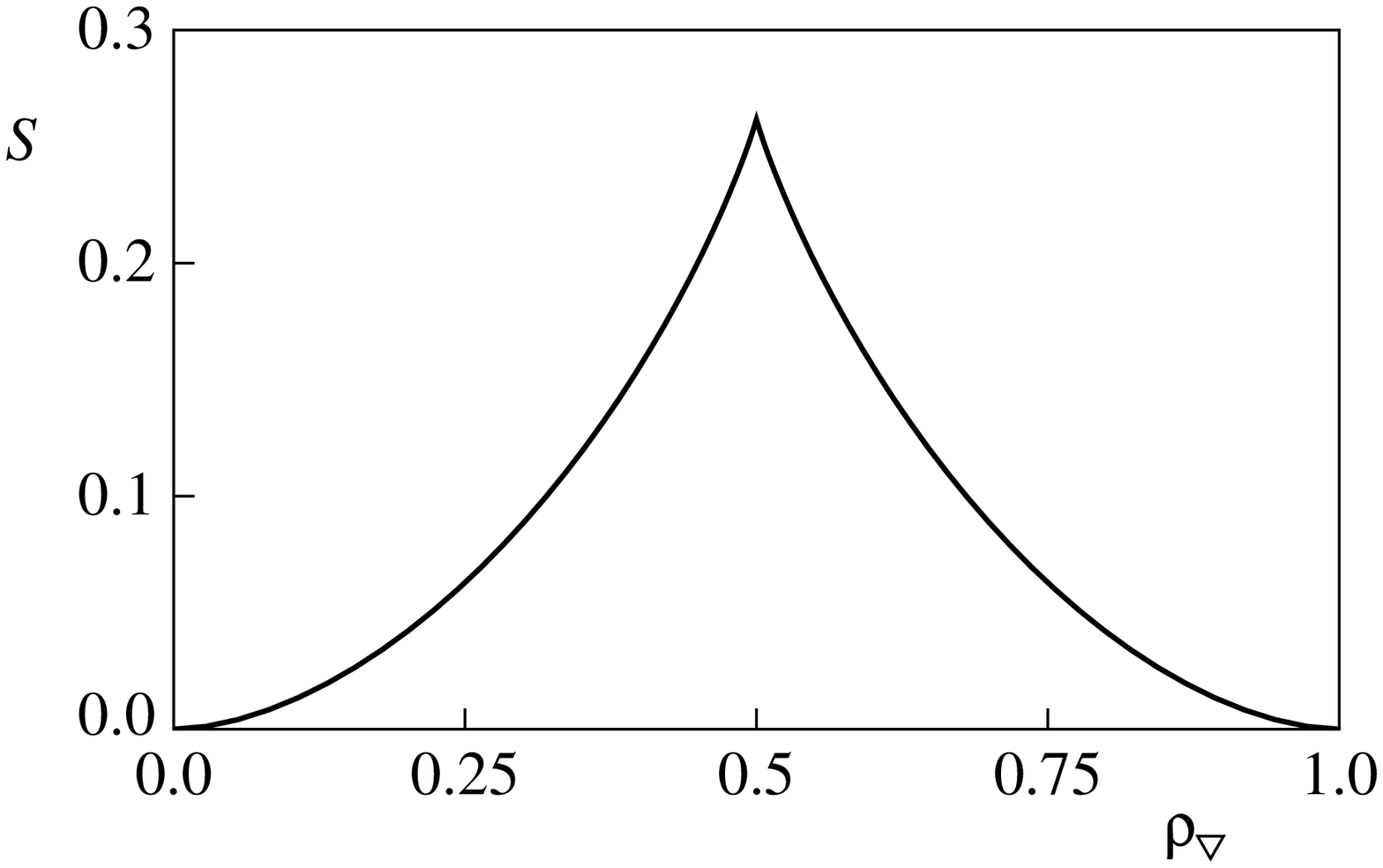,width=0.9 \columnwidth}\hfil
  \caption{The entropy per trimer versus the density
    $\rho_{\triangledown}$ of down trimers in the special case that
    $\rho_1 = \rho_3 = \rho_5$ and $\rho_2 = \rho_4$ for
    $\rho_{\triangledown} \le 1/2$, and similarly for
    $\rho_{\triangledown} \ge 1/2$.}
  \label{fig:entr}
\end{figure}

In the limit $\hat b \to 0$ the domain wall densities
$\rho_{\text{L}} = \rho_{\text{R}}$ vanish; the system is then filled
with sub-lattice~0 trimers and its entropy is zero.  At
$\hat b = 2 \text{i}$ the system is in a symmetric phase with all
sub-lattice densities equal~to $1/6$; its entropy
is~$S_{\text{sym}} = \log (3 \sqrt{3}/4)$.  When $\hat b$ is taken
between $0$ and $2 \text{i}$ on the imaginary axis, $b_{\text{L}}$ and
$b_{\text{R}}$ lie on the unit circle.  Then
$\rho_{\text{L}} = \rho_{\text{R}}$, so the sub-lattice densities
satisfy $\rho_2 = \rho_4$.  For small $\rho_{\triangledown} \le 1/2$
this equation together with $\rho_1 = \rho_3 = \rho_5$ describes the
most symmetric case for the sub-lattice densities.  Based on numerical
Bethe Ansatz calculations we believe that for given
$\rho_{\triangledown} \le 1/2$ the system takes its maximum entropy at
these sub-lattice densities.  Fig.~\ref{fig:entr} shows the entropy $S$
as function of~$\rho_{\triangledown}$.  The entropy for
$\rho_{\triangledown} \ge 1/2$ was obtained from that for
$\rho_{\triangledown} \le 1/2$ using the symmetry between the up and
the down trimers.

The entropy $S$ is a convex function of $\rho_{\triangledown}$ for
$0 \le \rho_{\triangledown} \le 1/2$.  A system with
$\rho_{\triangledown}$ in this interval is thermodynamically unstable.
It would separate into a phase with $\rho_{\triangledown} = 0$ and a
phase with $\rho_{\triangledown} = 1/2$, except for the fact that the
model does not admit an interface between these two phases.  Similarly
a system with $1/2 \le \rho_{\triangledown} \le 1$ would demix into
phases with $\rho_{\triangledown} = 1/2$ and
$\rho_{\triangledown} = 1$.

The transition between these phases can also be controlled by assigning
a chemical potential $\mu$ to the down trimers instead of imposing
their density~$\rho_{\triangledown}$.  From Fig.~\ref{fig:entr} it is
seen that for $\mu \le -2 S_{\text{sym}}$ the free energy
$F = -\mu \rho_{\triangledown} - S(\rho_{\triangledown})$ takes its
minimum at $\rho_{\triangledown} = 0$, so all trimers are on one of the
up sub-lattices.  For $-2 S_{\text{sym}} \le \mu \le 2 S_{\text{sym}}$
the minimum of $F$ is at $\rho_{\triangledown} = 1/2$, so all
sub-lattices are equally occupied.  For $\mu \ge 2 S_{\text{sym}}$ the
minimum of $F$ is at $\rho_{\triangledown} = 1$, so the system is again
in a frozen phase.

We thank Jan de Gier for fruitful discussions.  This work is part of
the research programme of the ``Stichting voor Fundamenteel Onderzoek
der Materie (FOM)'', which is financially supported by the
``Nederlandse Organisatie voor Wetenschappelijk Onderzoek (NWO)''.

\end{document}